\documentclass[pra,twocolumn,showpacs,floatfix,a4paper]{revtex4}%
\usepackage{graphics}
\usepackage{amsmath}
\usepackage{graphicx}
\usepackage{amsfonts}
\usepackage{amssymb}
\usepackage{float}
\usepackage{longtable}
\usepackage{epsfig}
\usepackage{latexsym}
\usepackage{theorem}
\usepackage{bbm}
\usepackage{bm}
\usepackage{psfrag}%
\begin{document}
\title{General immunity and superadditivity of two-way Gaussian quantum cryptography}
\date{\today}

\begin{abstract}
We consider two-way continuous-variable quantum key distribution, studying its
security against general eavesdropping strategies. Assuming the asymptotic
limit of many signals exchanged, we prove that two-way Gaussian protocols are
immune to coherent attacks. More precisely we show the general superadditivity
of the two-way security thresholds, which are proven to be higher than the
corresponding one-way counterparts in all cases. We perform the security
analysis first reducing the general eavesdropping to a two-mode coherent
Gaussian attack, and then showing that the superadditivity is achieved by
exploiting the random on/off switching of the two-way quantum communication.
This allows the parties to choose the appropriate communication instances to
prepare the key, accordingly to the tomography of the quantum channel. The
random opening and closing of the circuit represents, in fact, an additional
degree of freedom allowing the parties to convert, a posteriori, the two-mode
correlations of the eavesdropping into noise. The eavesdropper is assumed to
have no access to the on/off switching and, indeed, cannot adapt her attack.
We explicitly prove that this mechanism enhances the security performance, no
matter if the eavesdropper performs collective or coherent attacks.

\end{abstract}

\author{Carlo Ottaviani}
\email{Carlo.Ottaviani@york.ac.uk}
\author{Stefano Pirandola}
\affiliation{Department of Computer Science \& York Centre for Quantum Technologies,
University of York, YO10 5GH, UK}
\maketitle

\section*{Introduction}

Quantum Key Distribution (QKD) \cite{BB84} is today one of the most advanced
quantum technologies among those emerged from the fundamental research in
quantum information. Rapidly progressing towards practical implementations
\cite{scarani}, the interest in QKD is motivated by the promise of achieving
efficient distribution of cryptographic keys over insecure channels. In fact
its main goal is to provide an information-theoretic secure strategy to share
cryptographic keys in order to replace the current computationally-secure
solution \cite{RSA}, which has been proved to be vulnerable \cite{SHOR2} to
attacks by quantum computers.

The typical scenario involves two parties, Alice and Bob, who want
to share a secret message over an insecure channel
\cite{GISIN-RMP}~. To achieve this goal they encode classical
information in non-orthogonal quantum states, which are sent over
a noisy quantum channel under control of an eavesdropper, Eve. The
standard assumptions to analyze the security of QKD protocols are
the following: Eve is computationally unbounded, but has no-access
to the parties' private spaces \cite{GISIN-RMP,scarani} and, most
importantly, she is restricted by the no-cloning theorem
\cite{nocloning}. The distribution of private keys is possible
because any attempt to extract the encoded information unavoidably
introduces noise on the quantum states. Monitoring this noise the
parties can quantify how much Eve has learnt on the secret key
and, consequently, apply classical error correction and privacy
amplification protocols reducing Eve's information to a negligible
amount. Once they have distilled such a key, the parties can
safely use the one-time pad protocol. In case the level of noise
is too high, above the security threshold, they can abort the
protocol (denial of service).

The first theoretical proposals for QKD protocols have been
designed for discrete variables (DV) \cite{BB84} systems. Today
several remarkable implementation of DV-QKD have been achieved in
both fibers \cite{ZBINDEN-NAT-PHOT} and free space
\cite{QKD144-Zeilinger}. Beside this approach, several protocols
exploiting quantum continuous-variable (CV) systems have been put
forward
\cite{grosshansCOH,grosshansEB,no-switching,patron-hom-het,grosshansNAT,pirs}.
In CV-QKD \cite{diamanti} the information is encoded in quantum
systems with continuous spectra (infinite-dimensional Hilbert
space), and a special attention has been devoted to Gaussian CV
systems \cite{CWRMP}.

Gaussian CV-QKD has been achieved in \emph{in-field} implementations
\cite{jouguetNATPHOT}, with practical performances comparable to those of
DV-QKD, despite the latter appears to be more robust for long-distances. The
result of Ref.~\cite{jouguetNATPHOT} has been possible combining efficient
reconciliation protocols \cite{grosshansEB}, post-selection \cite{Silberhorn}
techniques and efficient classical compression codes \cite{leverrier}. The
interest in optical CV systems, for quantum information purposes, is now
growing, boosted mainly by the natural properties of these systems: relatively
cheap experimental implementation, higher rates, broadband detection
techniques \cite{SAM-RMP}, and the possibility of exploiting a wide range of
frequencies \cite{thermal-PRL,thermal-PRA}. These natural properties make
CV-QKD a promising candidate for future practical real-world implementations,
especially in the mid-range distances like the metropolitan areas where high
rates are desirable \cite{MDI-COMPARISON}.

Today, many theoretical efforts are devoted to the design of device
independent (DI) QKD protocols \cite{EKERT,gisin-MDI}. Despite recent
remarkable results, the practical implementation of this approach remains
still difficult \cite{VAZIRANI,COLBECK,BELL-TEST}. Very likely the next
generation of end-to-end quantum networks will use the recently introduced
\cite{SidePRL,Lo} measurement device independent QKD (MDI-QKD) which allows
the distribution of cryptographic keys preserving the protection against the
most typical side-channel attacks, without the need to pass a Bell test (see
Ref.~\cite{SidePRL} for a general security analysis). Recently a very
high-rate CV-MDI QKD protocol has been proposed and tested in a proof of
principle experiment \cite{PIRS-RELAY,SYMMETRIC-RELAY,MDI-COMPARISON}.

Alongside the study of end-to-end QKD, it is also of great interest the design
of more robust \textit{point-to-point} QKD schemes improving the security
performances of CV-QKD in noisier environments \cite{pirs} or able to exploit
trusted noise~\cite{discord} to implement QKD at different frequencies
\cite{thermal-PRL}. In this regard, the two-way protocols~\cite{pirs}, where
the parties make a double use of the quantum channel to improve the tolerance
to noise, show higher security thresholds than the one-way counterparts. This
idea has been also extended to thermal QKD \cite{thermal-2-PRA}. Also note
that the two-way protocols have been developed for DV-QKD \cite{BOSS,LM05} and
can be used for direct quantum communication \cite{SHAPIRO1,SHAPIRO2}.

\begin{figure}[ptb]
\vspace{-0.0cm}
\par
\begin{center}
\includegraphics[width=0.45\textwidth]{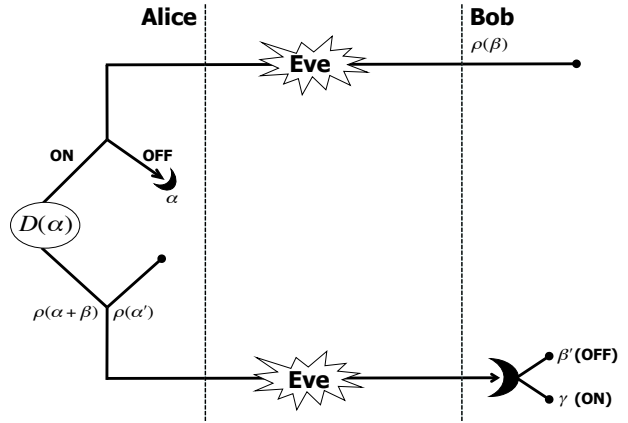}
\end{center}
\par
\vspace{-0.0cm}\caption[Two-way general scheme]{Two-way CV-QKD protocol.
Steps: \textit{forward}, Bob prepares coherent states of amplitude $\beta$ and
density matrix $\rho(\beta)$ and sends them through the noisy channel. Using
the circuit in ON configuration, Alice applies a random displacement
$D(\alpha)$ on $\rho(\beta)$ encoding information in the amplitude $\alpha$.
\textit{Backward}, Alice then sends the quantum state $\rho(\alpha+\beta)$ to
Bob who applies heterodyne detection with outcome $\gamma\simeq\alpha+\beta$,
and applies classical post-processing to subtract the reference amplitude
$\beta$ to recover $\alpha$. In OFF configuration, the circuit is opened at
Alice's station. She applies heterodyne detection on the reference state,
obtaining the variable $\alpha$. She then prepares a new coherent state
$\rho(\alpha^{\prime})$ which is sent back to Bob who heterodynes this state
obtaining the variable $\beta^{\prime}$. During the quantum communication Eve,
as well as Bob, does not know which setup of the circuit has been adopted. For
this reason, she is forced to attack both communication steps, and cannot
adapt her attack to the ON/OFF setup.}%
\label{2wayscheme}%
\end{figure}

The main result in this work is the explicit study of the asymptotic security
of two-way Gaussian protocols against coherent attacks, and the proof that
these schemes are in fact immune to this eavesdropping. The general strategy
to achieve this goal follows a previous insight \cite{pirs} and can be
summarized as follows (see Fig.~\ref{2wayscheme}): The parties randomly switch
ON or OFF the two-way communication line, and they post-select the
OFF\ instances if they detect the presence of coherent attacks, otherwise they
use the ON instances. We explicitly study the security threshold of the OFF
configuration against two-mode coherent attacks, which are the residual
eavesdropping after de Finetti reduction. Our approach allows us to prove that
the superadditivity of the two-way thresholds is a general feature. This
result can also be understood noting that the ON/OFF switching activates an
additional degree of freedom, exclusive to the parties, which can be used to
convert (a-posteriori) Eve's correlations into a noise on which Eve has no
control. 

\section*{Results}

\subsection*{The Scheme\label{SCHEME}}

In Fig.~\ref{2wayscheme} we describe a two-way quantum communication protocol.
We focus on use of coherent states, for the encoding, and heterodyne
detections for the decoding \cite{no-switching,pirs}. Bob prepares a
Gaussian-modulated reference coherent state with density matrix $\rho(\beta)$,
and use the quantum channel to transmit it to Alice who, randomly, decides to
close (case ON) or open (case OFF) the quantum communication. Let discuss the
two cases in detail.

\textit{Case ON}: Alice encoding is performed by applying a Gaussian-modulated
displacement $\hat{D}(\alpha)$ on the reference state $\rho(\beta)$,
obtaining\ a new coherent state with density matrix $\rho(\alpha+\beta)$. This
is sent back to Bob who performs heterodyne detection on the received state
$\rho(\alpha+\beta)$, and applies classical post-processing in order to
subtract the reference variable $\beta$ and derive Bob's estimate
$\tilde{\beta}$ of Alice's variable $\alpha$.

\textit{Case OFF}. Alice applies heterodyne detection on the reference state
$\rho(\beta)$ with outcome $\alpha$. Then, she prepares a new
Gaussian-modulated coherent state $\rho(\alpha^{\prime})$ which is sent back
to Bob, who heterodynes it with outcome $\beta^{\prime}$. After this, the
parties can use the two pairs of variables $\{\alpha,\beta\}$ and
$\{\alpha^{\prime},\beta^{\prime}\}$ to prepare the key.

We note that, during the quantum communication, both Bob and Eve do not know
the configuration adopted. This information is shared during the phase of
parameter estimation and is part of the classical communication performed by
Alice over the public channel. In the following we focus on the use of reverse
reconciliation (RR)~\cite{CWRMP,grosshansEB} (direct reconciliation is
discussed in the supplemental material). With the quantum communication in ON,
the RR corresponds to Alice inferring Bob's final outcome variable
$\tilde{\beta}$. With the circuit in OFF, the RR corresponds to Bob estimating
Alice's variable $\alpha$ during the forward stage, followed by Alice
estimating Bob's detection variable $\beta^{\prime}$.

As described in Ref.~\cite{pirs} the advantage of having the ON/OFF switching
is that this degree of freedom can be used to post-select the data in order to
prepare the key. After the channel tomography they can determine which attack
has been performed and in which status of the circuit. Then they keep data
from case ON when they detect a collective attack, while they use data
exchanged with the circuit in OFF when the attack is coherent.

\subsection*{Security analysis and attack reduction}

We study the security of the scheme assuming the asymptotic limit of many uses
of the quantum channel, $N\gg1$. In the worst-case scenario the eavesdropper
attaches ancillary quantum systems, $E_{k},$ to each exchanged signal, and
process the $E_{k}$'s by a global coherent unitary operation. The ancillary
output modes are stored in a quantum memory (QM), and coherently measured
after the classical communication between Alice and Bob at the end of the
protocol. Such an eavesdropping defines a general coherent attack.

The parties can now reduce the complexity of the previous scenario, by
applying symmetric random permutations \cite{renner-cirac} of their classical
data. This allows them to get rid of all the correlations between distinct
instances of the protocol. It is then clear that, in the case of two-way
communication, the de Finetti symmetrization provides a residual two-mode
coherent attack, where the only surviving correlations are those between
$E_{1}$ and $E_{2}$, used by Eve in each single round-trip. These ancillary
modes are mixed with the forward and backward signals by means of beam
splitters. Note that we can rid of additional modes $\mathbf{e}$ because we
work in the asymptotic limit and we bound Eve's accessible information using
the Holevo function \cite{pirsPRLATTACKS}. Finally, a further simplification
comes from the extremality of Gaussian states \cite{CWRMP}, which means that
we can restrict Eve's input $\rho_{E_{1}E_{2}}$ to be a Gaussian state.

\begin{figure*}[t]
\vspace{-0.0cm}
\par
\begin{center}
\includegraphics[width=0.75\textwidth]{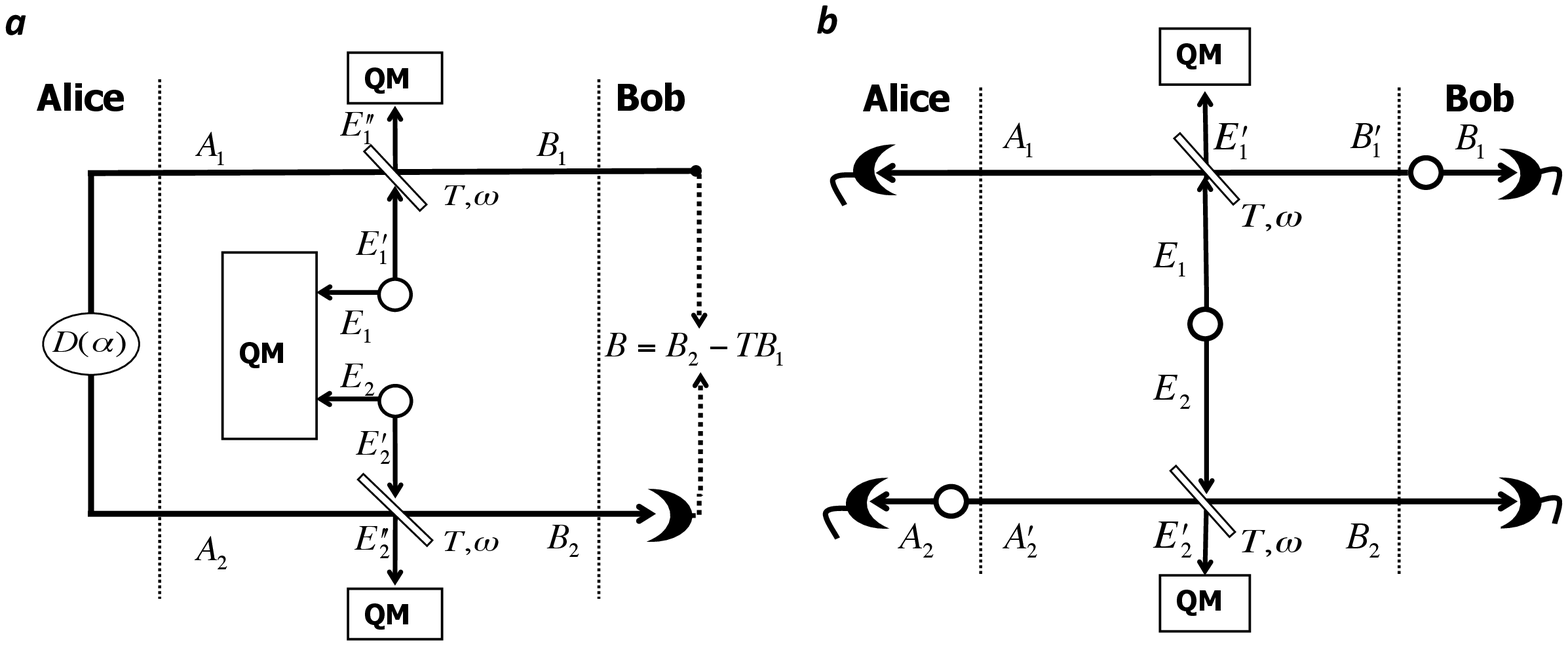}
\end{center}
\par
\vspace{-0.0cm}\caption[Two-way ON and OFF protocols]{Panel (a) shows the
two-way scheme in ON configuration. This is used against collective Gaussian
attacks, typically implemented by means of two independent entangling cloners.
Each beam-splitter has transmissivity $T$, and Eve mixes the ancillas
$E_{1}^{\prime}$ and $E_{2}^{\prime}$ with the signals modes $B_{1}$ and
$A_{2}$ Panel (b) describes the OFF configuration, whose security is studied
against two-mode Gaussian attacks. In this case we study the security of the
scheme in the entanglement based representation.}%
\label{fig:2wayONandOFF}%
\end{figure*}

The Gaussian attack is collective, when $E_{1}$, $E_{2}$ are uncorrelated, or
two-mode coherent when they are correlated. Studying this second case with the
circuit in ON and in DR, Ref.~\cite{2way-VS-COH} found that optimal two-mode
attacks exist which can reduce the security performances of the two-way
protocol below the one-way threshold. Here we show that, using the scheme with
the ON/OFF switching, and post-selecting the optimal key-rate accordingly to
the attack detected, the parties can overcome this problem. The security
analysis is performed according to the setup shown in
Fig.~\ref{fig:2wayONandOFF}, where Fig.~\ref{fig:2wayONandOFF}(a) refers to
collective attacks, while Fig.~\ref{fig:2wayONandOFF}(b) refers to two-mode
coherent attacks. In the latter case, the security analysis is performed in
the entanglement based (EB) representation \cite{grosshansEB,CWRMP}.

\subsubsection*{Description of the two-mode Gaussian attack in the EB
representation}

In EB representation both Bob and Alice remotely prepare coherent
states on the travelling modes $B_{1}^{\prime}$ and
$A_{2}^{\prime}$ by using two-mode squeezed vacuum states,
described by covariance matrices (CMs) of the
following form%
\begin{align}
V_{B_{1}B_{1}^{\prime}}  &  =\left(
\begin{array}
[c]{cc}%
\mu_{B}\mathbf{I} & \sqrt{\mu_{B}^{2}-1}\mathbf{Z}\\
\sqrt{\mu_{B}^{2}-1}\mathbf{Z} & \mu_{B}\mathbf{I}%
\end{array}
\right)  ,\label{BOB-EPR}\\
V_{A_{2}A_{2}^{\prime}}  &  =\left(
\begin{array}
[c]{cc}%
\mu_{A}\mathbf{I} & \sqrt{\mu_{A}^{2}-1}\mathbf{Z}\\
\sqrt{\mu_{A}^{2}-1}\mathbf{Z} & \mu_{A}\mathbf{I}%
\end{array}
\right)  , \label{ALICE-EPR}%
\end{align}
on which they apply heterodyne detections on the respective local
modes $B_{1}$ and $A_{2}$. The parameters $\mu_{A}=\mu+1$, and
$\mu_{B}=\mu+1$ describe the variance of the thermal state
injected by Alice and Bob, respectively. The two travelling modes,
$B_{1}^{\prime}$ and $A_{2}^{\prime}$, are mixed with Eve's modes,
$E_{1}$ and $E_{2}$, on two identical beam splitters, with
transmissivity $T$. Eve's input state $\rho_{E_{1}E_{2}}$ is a
zero mean, two-mode correlated thermal state, with CM%
\begin{equation}
V_{E_{1}E_{2}}=\left(
\begin{array}
[c]{cc}%
\omega\mathbf{I} & \mathbf{G}\\
\mathbf{G} & \omega\mathbf{I}%
\end{array}
\right)  ,\text{ with }\mathbf{G=}\left(
\begin{array}
[c]{cc}%
g & \\
& g^{\prime}%
\end{array}
\right)  , \label{CMattack}%
\end{equation}
where $\omega\geq1$ gives the variance of the thermal noise injected, while
$g$ and $g^{\prime}$ describe the correlations between the two ancillas.

Note that the double use of the channel corresponds to a sequential use of the
same communication line (optical fibre), so it is reasonable to consider a
symmetric channel ($T$ and $\omega$ are the same during the forward and
backward communication). The correlation parameters $g$ and $g^{\prime}$ must
fulfill the following constraints%
\begin{equation}
|g|<\omega,~|g^{\prime}|<\omega,~\omega^{2}+gg^{\prime}-1\geq\omega\left\vert
g+g^{\prime}\right\vert , \label{CMconstraints}%
\end{equation}
in order to certify that $V_{E_{1}E_{2}}$ is a bona fide CM. If $g^{\prime}=g
$, we must have $1-\omega\leq g\leq\omega-1$, with the two extremal conditions
corresponding to $E_{1}$ and $E_{2}$ sharing maximal separable correlations
\cite{pirs-DISC,pirs-sera}. If $g^{\prime}=-g$ the ancillas share
non-separable correlations. The Eq.~(\ref{CMconstraints}) provides the bound
$-\sqrt{\omega^{2}-1}\leq g\leq\sqrt{\omega^{2}-1}$, with the extremal values
corresponding to maximally entangled states. Finally, if $g=g^{\prime}=0$, the
two ancillas are not correlated, and the two-mode attack collapses to a
standard collective one, based on two independent entangling cloners.

\section*{Key-rates and security thresholds}

We compute now the secret-key rate $R:=I-I_{E}$, where $I$ is Alice-Bob mutual
information, and $I_{E}$ is Eve's accessible information. In the asymptotic
case $N\gg1$, $I_{E}$ can be replaced by the Holevo information $\chi$. Hence
we write%
\begin{equation}
R:=I-\chi. \label{KEY-RATE-WITH-HOLEVO}%
\end{equation}
The goal of the security analysis is the computation of the bound $\chi$,
which is defined as
\begin{equation}
\chi:=S_{E}-S_{E|\mathbf{x}}. \label{HOLEVO-DEF}%
\end{equation}
The functional $S_{E}$ is the von Neumann entropy, relative to Eve's quantum
state $\rho_{E}$, and $S_{E|\mathbf{x}}$ is that corresponding to
$\rho_{E|\mathbf{x}}$, which describes Eve's sate conditioned on the outcomes
of the measurements performed by the parties.

Against collective attacks, the parties use the protocol in ON, and we have
the following ON key-rate
\begin{equation}
R_{ON}:=I_{ON}(\alpha:\tilde{\beta})-\chi_{ON}(\varepsilon:\tilde{\beta}).
\label{KEY-RATE-DEF-ON}%
\end{equation}
By contrast, against coherent attacks, they use the circuit in OFF, for which
we have the following OFF key-rate
\begin{equation}
R_{OFF}=I_{OFF}-\chi_{OFF}, \label{KEY-RATE-DEF-OFF}%
\end{equation}
where%
\begin{equation}
I_{OFF}=\frac{I_{OFF}(\alpha:\beta)+I_{OFF}(\alpha^{\prime}:\beta^{\prime}%
)}{2} \label{IAB-OFF}%
\end{equation}
is the mutual information averaged over the forward and backward use, and%
\begin{equation}
\chi_{OFF}:=S_{AB}-S_{AB|\mathbf{\alpha,\beta}^{\prime}}\label{CHI-OFF-DEF}%
\end{equation}
is computed on Alice and Bob's output state $\rho_{AB}$.

Thus, for collective attacks we select the ON key-rate $R_{ON}$, while for
coherent attacks we select the OFF key-rate $R_{OFF}$. Both these key-rates
are function of channel parameters, i.e., transmissivity $T$ and excess noise
$N:=(1-T)(\omega-1)/T$ (which gives the extra noise on the channel with
respect the vacuum shot-noise). The OFF key-rate, $R_{OFF}$, is also function
of the correlation parameters $g$ and $g^{\prime}$. Therefore, once we have
$R$, we find the security thresholds solving the following equation
\begin{equation}
R(T,N,g,g^{\prime})=0. \label{SOGLIA}%
\end{equation}
This condition provides threshold curves of the type $N=N(T,g,g^{\prime})$
which simplifies to $N=N(T)$ for collective attacks.

\subsection*{Formulas for the key-rates}

The computation of the secret-key rates can be performed using the
mathematical tools described in Ref.~\cite{CWRMP}. From the knowledge of the
CM describing the total and conditional states, we can compute the von Neumann
entropies and finally the key rates. For the protocol with coherent states and
heterodyne detection we find the following key-rates%
\begin{align}
R_{ON}  &  =\log\frac{2T(1+T)}{e(1-T)(1+\Lambda)}+\sum_{i=1}^{3}h(\bar{\nu
}_{i})-2h(\omega),\label{rate}\\
R_{OFF}  &  =\log\frac{2T}{e(1-T)(1+\tilde{\Lambda})}+\sum_{j=\pm}\frac
{h(\bar{\nu}_{j}^{\prime})-h(\nu_{j})}{2}, \label{Rate-OFF}%
\end{align}
where
\[
h(x):=\frac{x+1}{2}\log\frac{x+1}{2}-\frac{x-1}{2}\log\frac{x-1}{2}.
\]

In the previous formulas, the symplectic eigenvalues $\bar{\nu}_{i}$ are
computed numerically and we define $\Lambda:=T^{2}+(1-T^{2})\omega$ and
$\tilde{\Lambda}:=T+(1-T)\omega$. It is of particular interest the OFF
key-rate of Eq.~(\ref{Rate-OFF}), from which we notice that one can recover
the one-way key-rate in the case of collective attacks ($g=g^{\prime}=0$). The
expressions of the total and conditional symplectic eigenvalues can be
computed analytically for large modulation, being equal to
\begin{align}
\nu_{\pm}  &  \rightarrow\sqrt{(\omega\pm g)(\omega\pm g^{\prime}%
)},\label{spectrum-HET2-TOT-OFF}\\
\bar{\nu}_{\pm}^{\prime}  &  \rightarrow\frac{\sqrt{\lbrack\lambda_{\pm
}+1-T][\lambda_{\pm}^{\prime}+1-T)]}}{T}, \label{spectrum-HET2-COND-RR-OFF}%
\end{align}
where $\lambda_{\pm}=T+(\omega\pm g)(1-T)$ and $\lambda_{\pm}^{\prime
}=T+(\omega\pm g^{\prime})(1-T)$.

\subsection*{Protocol with coherent states and homodyne detection}

Here we give the key-rate $\tilde{R}$ for the two-way protocol with coherent
states and homodyne detection. The only change with respect to the previous
scheme is clearly the use of homodyne detection for decoding. With the circuit
in ON, Bob prepares coherent states, randomly displaced by Alice and finally
homodyned by Bob. With the protocol in OFF, Bob prepares coherent states and
Alice performs homodyne detection. Then she sends back newly prepared coherent
states which are homodyned by Bob. After some algebra, we obtain the following
analytical expressions for the key-rates
\begin{align}
\tilde{R}_{ON}  &  =\frac{1}{2}\log\frac{T^{2}+\omega+T^{3}(\omega
-1)}{(1-T)\Lambda}+h(\tilde{\nu})-h(\omega),\label{Rate-HET-HOM-RR-ON}\\
\tilde{R}_{OFF}  &  =\frac{1}{2}\log\frac{\sqrt[4]{(\omega^{2}-g^{2}%
)(\omega^{2}-g^{\prime2})}}{(1-T)\tilde{\Lambda}}-\sum_{i=\pm}\frac{h(\nu
_{i})}{2}. \label{Rate-HET-HOM-RR-OFF}%
\end{align}

In the ON key-rate of Eq.~(\ref{Rate-HET-HOM-RR-ON}), used against collective
attacks, the asymptotic symplectic eigenvalue $\tilde{\nu}$ can be computed
analytically as
\[
\tilde{\nu}:=\sqrt{\frac{\omega\lbrack1+T^{2}\omega(1-T)+T^{3}]}{T^{2}%
+\omega+T^{3}(\omega-1)}}.
\]
By contrast, the OFF key-rate of Eq.~(\ref{Rate-HET-HOM-RR-OFF}) is exploited
under coherent attacks, and the total symplectic eigenvalues $\nu_{i}$ are the
same as given in Eq.~(\ref{spectrum-HET2-TOT-OFF}). The details to obtain
Eqs.~(\ref{Rate-HET-HOM-RR-ON}) and (\ref{Rate-HET-HOM-RR-OFF}) can be found
in the supplemental material, where we also include the secret-key rates
computed in DR.

\section*{Discussion}

\begin{figure}[t]
\vspace{-0.0cm}
\par
\begin{center}
\includegraphics[width=0.45\textwidth]{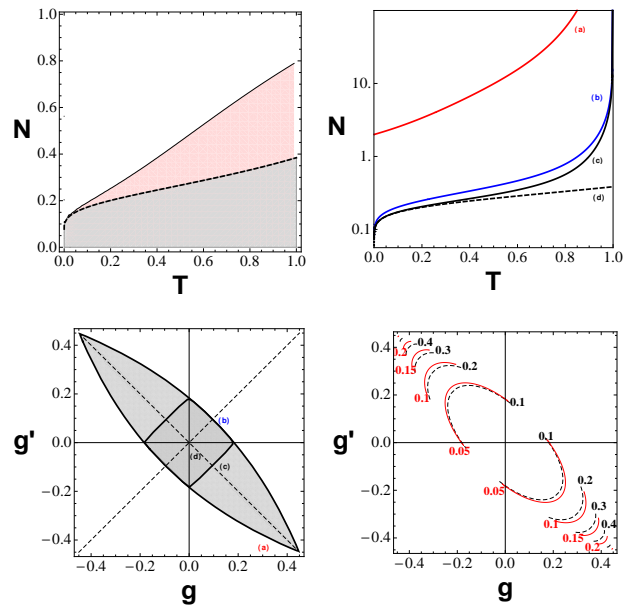}
\end{center}
\par
\vspace{-0.0cm}\caption[Two-way Vs. One-way ]{This figure summarizes the
results for the protocol with coherent states and heterodyne detection, whose
rates are given in Eqs.~(\ref{rate}) and (\ref{Rate-OFF}). The top panels
describe the security thresholds, in terms of tolerable excess noise $N$
versus transmissivity $T$. In the top-left panel, we consider collective
attacks and we compare the ON two-way threshold $R_{ON}=0$ (black solid line)
with the threshold of the one-way protocol (dashed line). In the pink region
the two-way protocol is secure, while the one-way counterpart is not. In the
top-right panel, we consider coherent attacks and we compare the OFF two-way
threshold (a)-(c) with respect to the one-way threshold (d). In particular,
curve (a) is obtained for $g=\pm\sqrt{\omega^{2}-1}$, i.e., Eve using
maximally entangled states; curve (b) considers the case $g^{\prime}=g$ with
$g=\pm(\omega-1)$; and curve (c) refers to $g^{\prime}=-g$ and $g=\pm
(\omega-1)$. Note that curve (d) coincides with the OFF threshold against
collective attacks, in which case the protocol is used in ON. The same labels
(a)-(d) are used in the bottom-left panel, which describes the various attacks
on the correlation plane $(g,g^{\prime})$, obtained setting $\omega
\simeq1.097$ in the constraint of Eq.~(\ref{CMconstraints}). Finally, in the
bottom-right panel, we plot the OFF key-rate against coherent attacks (red
lines), compared to the quantum mutual information (black lines) describing
the correlations of Eve's ancillas. We set $T=0.3$ and $\omega\simeq1.097$, so
that the one-way rate is $\simeq$ zero. We see that the OFF key-rate is always
strictly positive and it increases for increasing correlations in the attack.}%
\label{fig:2wayVs1way}%
\end{figure}

The security analysis of the thresholds coming from Eqs.~(\ref{rate}) and
(\ref{Rate-OFF}) is summarized in Fig.~\ref{fig:2wayVs1way}. In particular,
the security threshold for the ON configuration is confirmed~\cite{pirs} to be
superadditive in Fig.~\ref{fig:2wayVs1way}(top-left). The black-solid line
corresponds to the ON threshold, which is clearly above the threshold of the
one-way protocol (dashed line). This comparison is done against collective
attacks. The top-right panel of Fig.~\ref{fig:2wayVs1way}, shows the
security threshold for the OFF configuration in the presence of several
two-mode attacks with different values of the correlation parameters
$g=g^{\prime}$. The curves labeled (a)-(d) corresponds to the points in
Fig.~\ref{fig:2wayVs1way} (bottom-left). For example, the red curve (a)
describes coherent attacks performed with maximally entangled states. The
curve (b) describes attacks with $g^{\prime}=g=\pm(\omega-1)$, and the curve
(c) corresponds to $g^{\prime}=-g=\pm(\omega-1)$. Finally, the dashed curve
(d) corresponds to the center of the correlation plane, where the
OFF\ threshold coincides with the one-way threshold. Thus, we see that for any
value of Eve's correlation parameters, $g$ and $g^{\prime}$, Alice and Bob can
always post-select an instance of the two-way protocol (ON or OFF) whose
threshold is strictly greater than that of the corresponding one-way protocol.

Finally, Fig.~\ref{fig:2wayVs1way} (bottom-right) describes the connection
between the OFF key rate and the amount of correlations in Eve's ancillas, as
quantified by the quantum mutual information. We can see that the OFF key rate
not only is positive (while the one-way rate is always zero) but it also
increases with Eve's correlations, which are converted into noise by the OFF
configuration. Thus, the ON/OFF switching, together with the post-selection of
the correct instances, allows one to implement two-way CV-QKD in a way which
is not only secure, but also more robust to excess noise with respect to
one-way protocols under completely general attacks.

\section*{Methods}

A detailed description of the methods can be found in the supplementary
material. The security analysis of the protocol has been performed in the
entanglement based representation for the case OFF, so that we could compute
the Holevo bound from the study of Alice-Bob CM. For the case ON in RR, we
started from the output covariance matrix of Eve, to compute the total von
Neumann entropy. We then computed the conditional von Neumann entropy
completing Eve's covariance matrix with the Bob's mode on which we applied the
heterodyne or the homodyne measurement, accordingly with the case studied.

\section*{Conclusions}

In this work we have studied the security of two-way CV-QKD addressing,
explicitly, the superadditivity of its security threshold against coherent
attack. To the best of our knowledge this represents the first attempt of such
a complete study for direct point-to-point two-way protocols. Our security
analysis is obtained assuming the asymptotic limit, i.e., large number of
signals exchanged and large modulation. This allowed us to find closed
formulas for the secret-key rates, from which we have proved that the two-way
Gaussian protocols are more robust to excess noise than their one-way
counterparts in both collective and coherent attacks.

For this property, it is crucial the random ON/OFF switching of
the protocol, so that the eavesdropper's correlations are under
control of the parties and they are transformed, if needed, into
useless noise. Our analysis contributes to the general
understanding of the security properties of two-way protocols and
is useful to extend CV-QKD\ to regime with high excess noise.
Future developments could involve the study of this ON/OFF
switching strategy in more complex quantum communication
scenarios.

\section*{Acknowledgments}

We acknowledge financial support from the EPSRC via the `UK
Quantum Communications HUB' (Grant no. EP/M013472/1).

\newpage

\setcounter{section}{0} \setcounter{subsection}{0} \setcounter{figure}{0}
\setcounter{equation}{0} \renewcommand{\thefigure}{S\arabic{figure}}
\renewcommand{\figurename}{Figure}
\renewcommand{\theequation}{S\arabic{equation}} \renewcommand{\citenumfont}[1]{S#1}


\begin{center}
{Supplemental Material for \textquotedblleft General immunity and
superadditivity of two-way Gaussian quantum cryptography\textquotedblright}
\bigskip
\end{center}

This supplemental material gives the details of the calculations for the
security analysis of the protocols described in this work, and includes also
the key-rates calculated for two-way protocol used in direct reconciliation
(DR), which has not been discussed in the main body. We underline that
compared to the one-way protocol, the two-way presents a richer number of
cases which need to be analyzed. The protocols are named with respect to the
\textit{preparation} and \textit{detection} scheme adopted. Here we discuss
the protocol with coherent states and heterodyne detection, and with coherent
states and homodyne detection. Each one of previous cases can be implemented
in DR as well as reverse reconciliation (RR), and here we give the results for
both reconciliation schemes.

\subsection*{Secret-Key Rate and symplectic analysis}

The secret-key rate quantifies the gap between Alice and Bob's mutual
information and the information shared between Eve and the parties. Which
parties' variable(s) has(have) to be considered depends on the setup of the
protocol (one-way, two-way, ON or OFF) and, in general, from the
reconciliation protocol employed.

For instance consider the one-way protocol. We assume that Alice sends a
modulated coherent state with amplitude $\alpha$ to Bob, who receives a noisy
version of this state, whose amplitude is $\beta$. The parties can then obtain
two distinct secret-key rates defined as follows
\begin{align}
R^{\blacktriangleright}  &  :=I(\alpha:\beta)-\chi(\varepsilon:\alpha
),\label{rate-DR-sup-mat}\\
R^{\blacktriangleleft}  &  :=I(\alpha:\beta)-\chi(\varepsilon:\beta).
\label{rate-RR-sup-mat}%
\end{align}
The first describes the key-rate in DR, while the second the RR. The function
$I$ is the classical mutual information quantifying correlations between
Alice's variable, $\alpha$, and Bob's variable, $\beta$. For each quadrature
measured, and used to encode information, the mutual information is given by
the following general signal-to-noise ratio%
\begin{equation}
I=\frac{1}{2}\log\frac{V}{V_{C}}, \label{MUTUALINDEF}%
\end{equation}
where $V$ is the variance of the variable used to prepare the key, and $V_{C}
$ the conditioned variance of this statistical variable after the measurement
performed by the parties.

In the asymptotic limit of many uses of the quantum channel we can bound Eve's
accessible information by the Holevo function, which is given by
\begin{equation}
\chi(\varepsilon:x):=S(\varepsilon)-S(\varepsilon|x). \label{chi-supp-mat}%
\end{equation}
The function $S(.)$ describes the von Neumann entropy which, for Gaussian
quantum systems, has a simple form given by
\begin{equation}
S=\sum_{k}h(\nu_{k}), \label{vonNeumann-DEF}%
\end{equation}
with the entropic function $h(.)$ is defined as follows%
\begin{equation}
h(\nu_{k}):=\frac{\nu_{k}+1}{2}\log\frac{\nu_{k}+1}{2}-\frac{\nu_{k}-1}{2}%
\log\frac{\nu_{k}-1}{2}, \label{H-FUNC}%
\end{equation}
and where the $\nu_{k}$'s are the symplectic eigenvalues of the CM which
describes the dynamics of the studied Gaussian quantum system \cite{RMP}.

The expression of the von Neumann entropy of Eq.~(\ref{H-FUNC}) can be further
simplified exploiting the limit of large signal modulation, in which case we
can write \cite{RMP}
\begin{equation}
h(\nu_{k})=\log\frac{e}{2}\nu_{k}+O\left(  \frac{1}{\nu_{k}}\right)  .
\label{vonNeumann-DEF-asy}%
\end{equation}
The computation of the symplectic spectra can be done in prepare and measure
configuration, in which case the $\nu_{k}$'s are obtained from the symplectic
analysis of Eve's output CM or, in case we use the equivalent EB
representation, from Alice-Bob's CM. This second approach is used in the
following, to study the OFF configurations, i.e., when we consider coherent attacks.

To compute the symplectic spectrum, we first compute the appropriate CM
$\mathbf{V}$ and then, from matrix
\[
\mathbf{M}=i \Omega\mathbf{V},
\]
where $\Omega=\oplus_{i=1}^{n}\mathbf{\tilde{\omega}}_{i},$ with
$\mathbf{\tilde{\omega}}_{i}$ the single-mode symplectic form given by
\[
\mathbf{\tilde{\omega}}_{i}=\left(
\begin{array}
[c]{cc}%
0 & 1\\
-1 & 0
\end{array}
\right)  ,
\]
we compute the ordinary eigenvalues, which come in pairs. The symplectic
spectrum is obtained taking their absolute value.

\subsection*{Protocol with coherent states and heterodyne detection}

We start showing how we obtain the ON key-rate for the protocol with coherent
states and heterodyne detection, which is described in Fig.~2~(a) of the main
text. The security analysis is performed using Eve's CM, obtained from the
outputs $\{E_{1},E_{1}^{\prime\prime},E_{2},E_{2}^{\prime\prime}\}$. From this
we obtain the total von Neumann entropy and, by simple conditioning procedure,
one can also compute Eve's conditional CM. This describes the conditional
state $\rho_{E_{1},E_{1}^{\prime\prime},E_{2} ,E_{2}^{\prime\prime}|\alpha}$,
for the protocol in DR. By contrast, to study the protocol in RR, we complete
Eve's output CM with Bob's post-processed output mode $B$, on which we apply
the heterodyne detection in order to obtain the conditional CM for this case.

\subsubsection*{Case ON}

Bob sends modulated coherent states to Alice providing, on average, a thermal
state with variance $\mu_{B}=\mu+1$, where $\mu$ accounts for the classical
Gaussian modulation on the top of the vacuum shot-noise. Alice applies an
additional random displacement, $D(\alpha)$, on the states received from Bob
with modulation variance $\mu_{ON}=\mu\ge0$.

\paragraph*{Mutual information}

Alice-Bob mutual information can be computed from the expression of the
variance of post-processed mode $\langle B^{2}\rangle$, given by
\begin{equation}
\langle B^{2}\rangle=[T^{2}+T\mu_{ON}+(1-T^{2})\omega]I, \label{B2}%
\end{equation}
from which, in the limit of large modulation $\mu_{ON}=\mu\rightarrow\infty$,
we obtain the signal variance
\begin{equation}
V=T\mu. \label{variance--BOB}%
\end{equation}
We then compute the conditional variance from Eq.~(\ref{B2}) by setting Alice
modulation $\mu_{ON}=0$, obtaining
\begin{equation}
V_{C}=T^{2}+(1-T^{2})\omega. \label{variance--BOB-conditional}%
\end{equation}
Finally, using Eqs.~(\ref{variance--BOB}) and (\ref{variance--BOB-conditional}%
) with the expression of the mutual information in case of heterodyne
detection
\begin{equation}
I:=\log\frac{V+1}{V_{C}+1}, \label{IAB-HET}%
\end{equation}
we obtain the following Alice-Bob mutual information in the limit of large
modulation
\begin{equation}
I_{ON}=\log\frac{T\mu}{1+T^{2}+(1-T^{2})\omega}. \label{IabHET2}%
\end{equation}

\paragraph*{Total Covariance Matrix}

We compute now the CM of Eve's output quantum state $\rho_{E_{1}^{\prime
},E_{1}^{\prime\prime},E_{2}^{\prime},E_{2}^{\prime\prime}}$. We arrange it in
the following normal form
\begin{equation}
\mathbf{V}_{E}=\left(
\begin{array}
[c]{cc}%
\mathbf{A} & \mathbf{C}\\
\mathbf{C}^{T} & \mathbf{B}%
\end{array}
\right)  , \label{VeveTOT}%
\end{equation}
where
\begin{align}
\mathbf{A}  &  \mathbf{:}=\left(
\begin{array}
[c]{cc}%
\omega\mathbf{I} & \sqrt{T(\omega^{2}-1)}\mathbf{Z}\\
\sqrt{T(\omega^{2}-1)}\mathbf{Z} & \Psi\mathbf{I}%
\end{array}
\right)  ,\nonumber\\
\mathbf{B}  &  \mathbf{:}=\left(
\begin{array}
[c]{cc}%
\omega\mathbf{I} & \sqrt{T(\omega^{2}-1)}\mathbf{Z}\\
\sqrt{T(\omega^{2}-1)}\mathbf{Z} & \tilde{\Psi}\mathbf{I}%
\end{array}
\right)  ,\nonumber\\
\mathbf{C}  &  \mathbf{:}=\left(
\begin{array}
[c]{cc}%
0 & \Xi\mathbf{Z}\\
0 & \Phi\mathbf{Z}%
\end{array}
\right)  ,
\end{align}
with
\begin{align}
\tilde{\Psi}  &  =[T\omega+(1-T)^{2}\omega+T(1-T)\mu_{B}]+(1-T)\mu
_{ON},\nonumber\\
\Psi &  =T(\omega-\mu_{B})+\mu_{B},\nonumber\\
\Phi &  =(1-T)\sqrt{T}(\mu_{B}-\omega),\nonumber\\
\Xi &  =-(1-T)\sqrt{(\omega^{2}-1)}. \label{mode-B}%
\end{align}

From Eq.~(\ref{VeveTOT}) we easily obtain the total symplectic spectrum by
taking the limits for $\mu_{ON}=\mu\rightarrow\infty$ and $\mu_{b}%
=\mu+1\rightarrow\infty$
\begin{equation}
\{\nu_{1},\nu_{2},\nu_{3}\nu_{4}\}\rightarrow\{\omega,\omega,(1-T)^{2}\mu
^{2}\}. \label{spectrum-het2-ON-DR}%
\end{equation}
The latter, used with Eqs.~(\ref{vonNeumann-DEF}) and
(\ref{vonNeumann-DEF-asy}), gives the total von Neumann entropy
\begin{equation}
S_{E}=\log\frac{e^{2}}{4}(1-T)^{2}\mu^{2}+2h(\omega) . \label{SE-HET2}%
\end{equation}

\paragraph*{Conditional CM and Key-rate in Direct Reconciliation}

For the DR the conditional CM can be obtained straightforwardly from
Eq.~(\ref{VeveTOT}) setting $\mu_{ON}=0$ on both quadratures in the block
describing Eve's output $E_{2}^{\prime\prime}$, i.e., $\tilde{\Psi}$ in
Eq.~(\ref{mode-B}). The resulting conditional CM has the following asymptotic
symplectic spectrum
\begin{equation}
\{\bar{\nu}_{1},\bar{\nu}_{2},\bar{\nu}_{3},\bar{\nu}_{4}\}\rightarrow
\{1,1,\omega,(1-T^{2})\mu\}. \label{spectrumCOND-ON-HET2-DR}%
\end{equation}
Using Eq.~(\ref{spectrumCOND-ON-HET2-DR}) with Eq.~(\ref{vonNeumann-DEF}) and
(\ref{vonNeumann-DEF-asy}), we compute the conditional von Neumann entropy
\begin{equation}
S_{E|\alpha}=\log\frac{e}{2}(1-T^{2})\mu+h(\omega). \label{SE-COND-HET2}%
\end{equation}
Now, using Eqs.~(\ref{SE-HET2}) and (\ref{SE-COND-HET2}) in
Eq.~(\ref{chi-supp-mat}), one obtains the expression of the Holevo bound
\begin{equation}
\chi_{ON}^{\blacktriangleright}=\log\frac{e}{2}\frac{(1-T)}{(1+T)}\mu.
\label{CHI-DR-ON-HET2}%
\end{equation}
Finally, by subtracting the Holevo function of Eq.~(\ref{CHI-DR-ON-HET2}) from
the mutual information of Eq.~(\ref{IabHET2}) we get the ON key-rate in DR
\[
R_{ON}^{\blacktriangleright}=\log\frac{2}{e}\frac{T(1+T)}{(1-T)[1+T^{2}%
+(1-T^{2})\omega]}-h(\omega).
\]

\paragraph*{Reverse Reconciliation}

To study the security of the protocol in RR, we need to re-compute the
conditional von Neumann entropy for this case. We complete the CM of
Eq.~(\ref{VeveTOT}) adding the blocks describing Bob's output mode $B$ and its
correlations with the rest of Eve's modes. Then we apply a heterodyne
detection on $B$ obtain Eve's conditional CM after Bob's measurements. We then
write
\begin{equation}
\mathbf{V}^{\blacktriangleleft}=\left(
\begin{array}
[c]{cc}%
\mathbf{V}_{E} & \mathbf{\bar{C}}\\
\mathbf{\bar{C}}^{T} & \mathbf{\bar{B}}%
\end{array}
\right)  , \label{V-EVE-AND-BOB-ON-RR}%
\end{equation}
where
\begin{align*}
\mathbf{\bar{B}}  &  \mathbf{=}[T^{2}+T\mu+(1-T^{2})\omega]\mathbf{I,}\\
\mathbf{\bar{C}}  &  \mathbf{=}\sqrt{1-T}\left(
\begin{array}
[c]{c}%
\sqrt{T(\omega^{2}-1)}\mathbf{Z}\\
T(\omega-1)\mathbf{I}\\
\sqrt{(\omega^{2}-1)}\mathbf{Z}\\
\sqrt{T}[T(\omega-1)-\mu]\mathbf{I}%
\end{array}
\right)  .
\end{align*}
We then apply the formula for heterodyne detection \cite{RMP} obtaining the
following conditional CM
\begin{equation}
\mathbf{V}_{C}^{\blacktriangleleft}=\mathbf{V}_{E}+\mathbf{\bar{C}(\bar{B}%
+I)}^{-1}\mathbf{\bar{C}}^{T}, \label{Veve-C-RR}%
\end{equation}
which gives the following conditional symplectic spectrum
\begin{equation}
\{\bar{\nu}_{1},\bar{\nu}_{2},\bar{\nu}_{3},\bar{\nu}_{4}\}\rightarrow
\{\bar{\nu}_{1},\bar{\nu}_{2},\bar{\nu}_{3},(1-T^{2})\mu\}.
\label{SPECTRUM-COND-HET2-ON}%
\end{equation}
Notice that the eigenvalues $\bar{\nu}_{1},\bar{\nu}_{2},\bar{\nu}_{3}$ are
asymptotically depending only on the channel parameters $(\omega, T)$, and are
related by the following expression
\[
\bar{\nu}_{1}\bar{\nu}_{2}\bar{\nu}_{3}=\frac{[1+T^{3}+(1-T)(1+T^{2}%
)\omega]\omega}{T(1+T)}.
\]
From the eigenvalues of Eq.~(\ref{SPECTRUM-COND-HET2-ON}), used with
Eqs.~(\ref{chi-supp-mat}), (\ref{vonNeumann-DEF}) and
(\ref{vonNeumann-DEF-asy}) we obtain the Holevo bound for the RR
\[
\chi_{ON}^{\blacktriangleleft}=\log\frac{e}{2}(1-T^{2})\mu+h(\bar{\nu}%
_{1})+h(\bar{\nu}_{2})+h(\bar{\nu}_{3}),
\]
which, used with Eq.~(\ref{IabHET2}) in the definition of
Eq.~(\ref{KEY-RATE-DEF-ON}) in the main text, gives the secret-key rate for
the protocol used in RR given in Eq.~(\ref{rate}) of the main text.

\subsection*{Case OFF}

We now describe the details of the calculations for the protocol used in OFF,
as described in Fig.~2-(b) of the main text. In this case we perform the
security analysis considering two-mode coherent attacks, in the EB representation.

\subsubsection*{Total Covariance Matrix and von Neumann entropy}

Bob starts from a two-mode squeezed vacuum state, described by the CM of
Eq.~(\ref{BOB-EPR}) in the main text. Applying a local heterodyne detection on
mode $B_{1}$, he projects the traveling mode $B_{1}^{\prime}$ in a coherent
state. In the same way, Alice applies a local heterodyne detection on mode
$A_{2}$, projecting the traveling mode $A_{2}^{\prime}$ in a coherent state.
Finally, we assume that Eve injects the general Gaussian state described by
Eq.~(\ref{CMattack}) in the main text. Since the total state of Alice, Bob and
Eve is pure, we can reduce ourselves to compute Alice and Bob's state (having
the same entropy of Eve's). We then order Alice's and Bob's the output modes
as follows $\{B_{1},$ $A_{2},$ $A_{1},$ $B_{2}\}$, and obtain the following
expression
\begin{equation}
\mathbf{V}_{AB}^{OFF}=\left(
\begin{array}
[c]{cc}%
\mathbf{\tilde{A}} & \mathbf{\tilde{C}}\\
\mathbf{\tilde{C}}^{T} & \mathbf{\tilde{B}}%
\end{array}
\right)  , \label{VeveTOT-OFF-HET2}%
\end{equation}
where the matrix blocks have been defined as follows,%
\begin{align*}
\mathbf{\tilde{A}}  &  \mathbf{=}\left(
\begin{array}
[c]{ccc}%
\mu_{B}\mathbf{I} &  & \tilde{\delta}\mathbf{I}\\
& \mu_{A}\mathbf{I} & \\
\tilde{\delta}\mathbf{I} &  & \tilde{\tau}(\mu_{B})\mathbf{I}%
\end{array}
\right)  \mathbf{,}\\
\mathbf{\tilde{B}}  &  =\tilde{\tau}(\mu_{A})\mathbf{I},\\
\mathbf{\tilde{C}}  &  =\left(
\begin{array}
[c]{c}%
\mathbf{0}\\
\tilde{\delta}\mathbf{I}\\
(1-T)\mathbf{G}%
\end{array}
\right)  ,
\end{align*}
where
\[
\mathbf{G:=}\left(
\begin{array}
[c]{cc}%
g & \\
& g^{\prime}%
\end{array}
\right)  ,
\]
the coefficients $\widetilde{\delta}$ and $\tilde{\tau}(y)$ have been defined
as follows,
\begin{align}
\tilde{\delta}  &  :=\sqrt{T[\mu_{B}{}^{2}-1]},\nonumber\\
\tilde{\tau}(y)  &  :=(1-T)\omega+Ty. \label{TAU-TILDE}%
\end{align}

We compute the symplectic spectrum of CM (\ref{VeveTOT-OFF-HET2})\ and taking
the asymptotic limit, $\mu_{A}=\mu_{B}\rightarrow\infty$, we obtain the
following analytical expressions
\begin{equation}
\{\nu_{\pm},\nu_{3},\nu_{4}\}\rightarrow\{\sqrt{(\omega\pm g)(\omega\pm
g^{\prime})},(1-T)\mu,(1-T)\mu\}, \label{spectrumTOT-OFF}%
\end{equation}
which gives the total von Neumann entropy for the case OFF
\begin{equation}
S_{AB}=\log\left(  \frac{e}{2}\right)  ^{2}(1-T)^{2}\mu^{2}+h(\nu_{-}
)+h(\nu_{+}). \label{vonNeumann-TOT-HET2-OFF}%
\end{equation}

\subsubsection*{Conditional covariance matrix and Alice-Bob mutual
information}

To obtain the conditional CM in DR we set $\mu_{A}=\mu_{B}=1$, in modes
$B_{1}$ and $A_{2}$. It is easy to verify that the resulting CM has the
following symplectic spectrum
\begin{equation}
\{\bar{\nu}_{1},\bar{\nu}_{2},\bar{\nu}_{+},\bar{\nu}_{-}\}\rightarrow
\{1,1,\sqrt{\lambda_{+}\lambda_{+}^{\prime}},\sqrt{\lambda_{-}\lambda
_{-}^{\prime}}\},\label{spectrumCOND-OFF-HET2-DR}%
\end{equation}
where, $\lambda_{\pm}=T+(\omega\pm g)(1-T)$ and $\lambda_{\pm}^{\prime
}=T+(\omega\pm g^{\prime})(1-T)$. Using these eigenvalues, we compute the
following conditional von Neumann entropy
\[
S_{AB|\alpha^{\prime},\beta}=h(\bar{\nu}_{+})+h(\bar{\nu}_{-}).
\]
The previous equation and Eq.~(\ref{vonNeumann-TOT-HET2-OFF}) are then used to
obtain the asymptotic expression of the Holevo function in DR, which is given
by
\begin{align}
\chi_{OFF}^{\blacktriangleright} &
:=\frac{S_{AB}-S_{AB|\mathbf{\alpha,\beta
}^{\prime}}}{2}.\\
& =\log\frac{e}{2}(1-T)\mu+\frac{1}{2}\sum_{k=\pm}\left[ h(\nu
_{k})-h(\bar{\nu}_{k})\right]  .\label{HOLEVO-OFF-HET2-DR}%
\end{align}

The conditional CM corresponding to the RR, is obtained by applying two
consecutive heterodyne detections, starting from CM of
Eq.~(\ref{VeveTOT-OFF-HET2}). We first measure mode $B_{2}$ and then we apply
another heterodyne detection on mode $A_{2}$ (the order of these two local
measurements is of course irrelevant). The resulting conditional CM has the
symplectic spectrum
\[
\bar{\nu}_{\pm}^{\prime}\rightarrow\frac{\sqrt{\lbrack\lambda_{\pm
}+1-T][\lambda_{\pm}^{\prime}+1-T)]}}{T}.
\]
These are used to compute the Holevo bound which is given by
\begin{equation}
\chi_{OFF}^{\blacktriangleleft}=\log\frac{e}{2}(1-T)\mu+\frac{1}%
{2}\sum_{k=\pm}[h(\nu_{k})-h(\bar{\nu}_{k}^{\prime}).
\label{HOLEVO-OFF-HET2-RR}%
\end{equation}

\subsubsection*{Alice-Bob Mutual Information and Secret-key rate.}

Alice-Bob mutual information is easily computed from the coefficient
$\tilde{\tau}$, given in Eq.~(\ref{TAU-TILDE}). Taking the limit of large
modulation, using the formula defining the mutual information for heterodyne
detections (\ref{IAB-HET}), and averaging over the double use of the quantum
channel we obtain the following expression
\begin{equation}
I_{OFF}=\log\frac{T\mu}{1+\tilde{\Lambda}}, \label{IabHET2-OFF}%
\end{equation}
where $\tilde{\Lambda}=T+(1-T)\omega.$ Notice that, differently from the
Holevo bound, the mutual information is independent from the correlation
parameters described by the matrix $\mathbf{G}$.

Using Eq.~(\ref{HOLEVO-OFF-HET2-DR}) and Eq.~(\ref{IabHET2-OFF}), and after
some simple algebra, we get the analytical expression of the key-rate in DR
\begin{equation}
R_{OFF}^{\blacktriangleright}=\log\frac{2T}{e(1-T)(1+\tilde{\Lambda})}+\frac{
1}{2}%
{\displaystyle\sum\limits_{k=\pm}}
[h(\bar{\nu}_{k})-h(\nu_{k})], \label{R-DR-HET2-OFF}%
\end{equation}
The key-rate in RR of Eq.~(\ref{Rate-OFF}) in the main text, is obtained using
previous Eq.~(\ref{HOLEVO-OFF-HET2-RR}) and Eq.~(\ref{IabHET2-OFF}).

\subsection*{Protocol with coherent states and homodyne detection}

\begin{figure}[t]
\vspace{-0.0cm}
\par
\begin{center}
\includegraphics[width=0.45\textwidth]{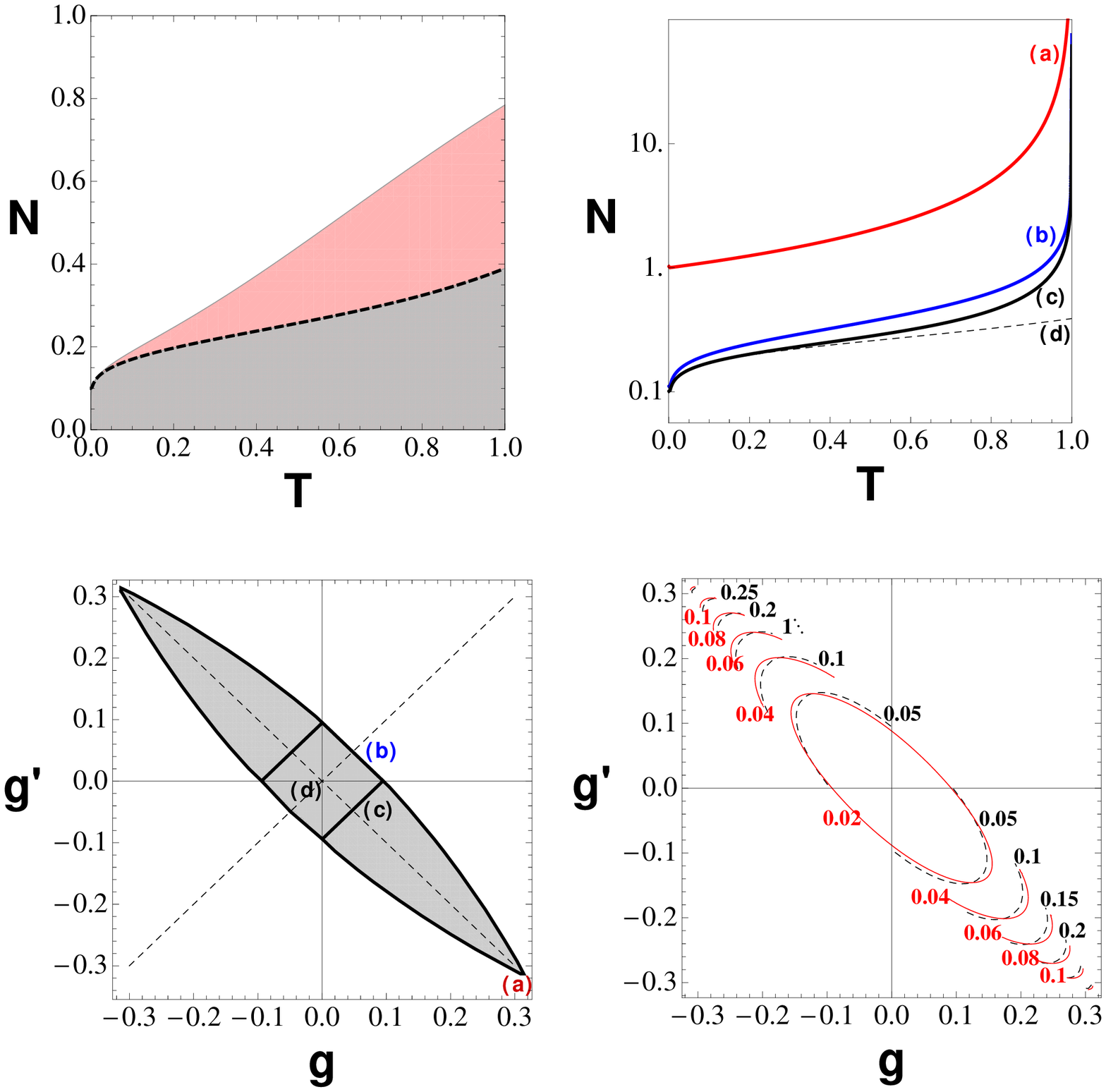}
\end{center}
\par
\vspace{-0.0cm}\caption[Two-way Vs. One-way coherent/homodyne protocol]{This
figure summarize the results for the protocol with coherent states and
homodyne detection, whose rates are given in Eqs.~(\ref{Rate-HET-HOM-RR-ON-SI}%
) and (\ref{Rate-HET-HOM-RR-OFF-SI}) which are the same of Eqs.~(16) and (17)
in the main text. The top panels give the security thresholds in terms of
tollerable excess noise $N$ versus transmissivity $T$. In the top-left panel,
we consider collective attacks and we compare the ON two-way threshold
$\tilde{R}_{ON}=0$ (black solid line) with the threshold of the one-way
protocol (dashed line). In the pink region the two-way protocol is secure,
while the one-way counterpart is not. In the top-right panel, we consider
coherent attacks and we compare the OFF two-way threshold (a)-(c) with respect
to the one-way threshold (d). In particular, curve (a) is obtained for
$g=\pm\sqrt{\omega^{2}-1}$, i.e., Eve using maximally entangled states; curve
(b) considers the case $g^{\prime}=g$ with $g=\pm(\omega-1)$; and curve (c)
refers to $g^{\prime}=-g$ and $g=\pm(\omega-1)$. Note that curve (d) coincides
with the OFF threshold against collective attacks, in which case the protocol
is used in ON. The same labels (a)-(d) are used in the bottom-left panel,
which describes the various attacks on the correlation plane $(g,g^{\prime})$,
obtained setting $\omega\simeq1.049$ in the constraint of
Eq.~(\ref{CMconstraints}) in the main text. Finally, in the bottom-right
panel, we plot the OFF key-rate against coherent attacks (red lines), compared
to the quantum mutual information (black lines) describing the correlations of
Eve's ancillas. We set $T=0.2$ and $\omega\simeq1.049 $, so that the one-way
rate is $\simeq$ zero. We see that the OFF key-rate is always strictly
positive and it increases for increasing correlations in the attack.}%
\label{fig:2wayVs1wayHETHOM}%
\end{figure}In contrast to the protocol analysed in previous section, here the
decodings are performed by means of homodyne detections. This modifies the
expression of the mutual information and these of the conditional von Neumann entropies.

\subsubsection*{Case ON}

\paragraph*{Direct Reconciliation: conditional covariance matrix}

The conditional CM in DR is obtained as before. We start from the total CM of
Eq.~(\ref{VeveTOT}) and we apply the following conditioning procedure
\begin{align}
\bar{\mu}_{ON}^{q}  &  =1/\mu\overset{\mu\rightarrow\infty}{\rightarrow
}0,\nonumber\\
\bar{\mu}_{ON}^{p}  &  =\mu, \label{conditioning-HOM}%
\end{align}
which describes Alice's effective modulation in order to describe the
measurement of only one quadrature during the decoding stage (homodyne detection).

\paragraph*{Direct Reconciliation: mutual information, Holevo bound and
key-rate}

The conditioning procedure, described by Eqs.~(\ref{conditioning-HOM}), can
clearly also used to determine Alice-Bob mutual information. In the present
case only one-quadrature is used to encode the key so Alice-Bob mutual
information is given by the following expression
\begin{equation}
\tilde{I}_{ON}=\frac{1}{2}\log\frac{T\mu}{T^{2}+(1-T^{2})\omega}.
\label{Iab-het-hom-ON}%
\end{equation}

Eve's conditional CM is obtained from Eq.~(\ref{VeveTOT}) applying recipe of
Eqs.~(\ref{conditioning-HOM}). One easily obtains the conditional symplectic
spectrum which, in the asymptotic limit, is given by
\begin{equation}
\{\nu_{1}^{\blacktriangleright},\nu_{2}^{\blacktriangleright},\nu
_{3}^{\blacktriangleright}\nu_{4}^{\blacktriangleright}\}\rightarrow
\{1,\omega,\sqrt{(1-T)^{2}(1-T^{2})\omega\mu^{3}}\}.
\label{spectrum-het-hom-ON-DR}%
\end{equation}
After some algebra we obtain the following Holevo bound
\[
\tilde{\chi}_{ON}^{\blacktriangleright}=\frac{1}{2}\log\frac{(1-T)^{2}\mu
}{(1-T^{2})\omega}+h(\omega),
\]
which, used with Eq.~(\ref{Iab-het-hom-ON}), provides the ON key-rate for the
protocol in DR
\begin{equation}
\tilde{R}_{ON}^{\blacktriangleright}=\frac{1}{2}\log\frac{T(1+T)\omega
}{(1-T)[T^{2}+(1-T^{2})\omega]}-h(\omega). \label{R-ON-HET-HOM-DR}%
\end{equation}
It is interesting to note that plotting the security threshold of the key-rate
of Eq.~(\ref{R-ON-HET-HOM-DR}), one finds that it provides a positive key-rate
even below $3$ dB, which sets the limit performance of the one-way version of
this protocol in DR.

\paragraph*{Reverse Reconciliation: conditional covariance matrix}

The security of the protocol is performed repeating the steps described in
previous sections, replacing the heterodyne detections with homodyne
measurements on $B$. Indeed, we apply the following formula
\[
\mathbf{\tilde{V}}_{C}^{\mathbf{\blacktriangleleft}}=\mathbf{A}-\mathbf{C}%
(\Pi\mathbf{\bar{B}}\Pi)\mathbf{C}^{T},
\]
to Eq.~(\ref{V-EVE-AND-BOB-ON-RR}). Note that $\Pi:=$diag$(1,0)$
$($diag$(0,1))$ for heterodyne on quadrature $\hat{q}$ $(\hat{p})$. We then
compute the symplectic spectrum of CM $\tilde{\mathbf{V}}_{C}%
^{\mathbf{\blacktriangleleft}}$, which we rewrite here in the following form
\begin{align}
\tilde{\nu}  &  \rightarrow\sqrt{\frac{\omega\lbrack1+T^{2}\omega-T^{3}%
(\omega-1)]}{T^{2}+\omega+T^{3}(\omega-1)}},\label{spectrum-het-hom-ON-RR}\\
\nu_{2}^{\blacktriangleleft}  &  \rightarrow\omega,\nonumber\\
\nu_{3}^{\blacktriangleleft}\nu_{4}^{\blacktriangleleft}  &  \rightarrow
\sqrt{\frac{(1-T)^{3}[T^{2}+\omega+T^{3}(\omega-1)]\mu^{3}}{T}}.\nonumber
\end{align}
We then obtain the Holevo bound $\tilde{\chi} _{ON}^{\blacktriangleleft}$
\[
\tilde{\chi}_{ON}^{\blacktriangleleft}=h(\omega)-h(\tilde{\nu})+\frac{1}%
{2}\log\frac{T(1-T)\mu}{T^{2}+\omega+T^{3}(\omega-1)},
\]
which combined with the mutual information of Eq.~(\ref{Iab-het-hom-ON}) gives
the ON key-rate in RR of Eq.~(\ref{Rate-HET-HOM-RR-ON}) of the main text,
i.e.,
\begin{align}
\tilde{R}_{ON}^{\blacktriangleleft}  &  =\frac{1}{2}\log\frac{T^{2}
+\omega+T^{3}(\omega-1)}{(1-T)\Lambda} + h\left(  \tilde{\nu}\right)
-h(\omega), \label{Rate-HET-HOM-RR-ON-SI}%
\end{align}
with $\Lambda:=T^{2}+(1-T^{2})\omega$.

\subsection*{Case OFF}

This case is studied in the EB representation, following the same steps of the
previous OFF case, for both the DR and RR, replacing the final heterodyne with
homodyne detections. The mutual information is computed averaging over the
double use of the quantum channel, i.e., using the following definition of the
mutual information
\begin{align}
\tilde{I}_{OFF}  & :=\frac{1}{2}\left(  \frac{1}{2}\sum_{i=A,B}\log\frac
{T\mu_{i}}{(1-T)\omega+T\mu_{i}}\right)  \nonumber\\
& \overset{\mu_{i}=\mu\rightarrow\infty}{=}\frac{1}{2}\log\frac{T\mu}%
{\tilde{\Lambda}},\label{I-OFF-HET-HOM}%
\end{align}
where $\tilde{\Lambda}:=T+(1-T)\omega$.

\subsubsection*{Direct Reconciliation}

The steps to compute the conditional CM have been discussed previously, so
here we just provide the analytical expressions of the conditional symplectic
spectra
\begin{align*}
\tilde{\nu}_{\pm}^{\blacktriangleright} &  =\sqrt{(1-T)\Gamma_{\pm}\mu},\\
\tilde{\eta}_{\pm}^{\blacktriangleright} &  =\sqrt{\frac{(\omega\pm g^{\prime
})[T+(\omega\pm g)(1-T)]}{\Gamma_{\pm}}},
\end{align*}
where we define
\[
\Gamma_{\pm}:=1-T+T(\omega\pm g^{\prime}).
\]
Notice that, in previous spectra, the role of the correlation parameters
depends on the quadrature measured by the homodyne detection of the decoding
stage. We obtain the following Holevo bound for the DR
\[
\tilde{\chi}_{OFF}^{\blacktriangleright}=\frac{1}{2}\log\frac{(1-T)\mu}%
{\sqrt{[1+t(\omega-1)]^{2}-T^{2}g^{2}}},
\]
and the key-rate in direct reconciliation is given by
\begin{align*}
\tilde{R}_{OFF}^{\blacktriangleright} &  =\frac{1}{2}\log\frac{T\sqrt
{[1+T(\omega-1)]^{2}-T^{2}g^{2}}}{(1-T)[T+(1-T)\omega]}\\
&  -{\sum\limits_{k=\pm}}\frac{h(\tilde{\eta}_{k}^{\blacktriangleright
})-h\left(  \nu_{k}\right)  }{2},
\end{align*}
where the eigenvalues $\nu_{\pm}$ are defined in Eq.~(\ref{spectrumTOT-OFF}).

\subsubsection*{Reverse Reconciliation}

For the RR, when the homodyne detection is performed on the quadrature
$\hat{q}$, we obtain the following conditional symplectic eigenvalues
\begin{equation}
\tilde{\nu}_{\pm}^{\blacktriangleleft}=\sqrt{\frac{(1-T)(\omega\pm g)\mu}{T}%
}.\label{SYMP-EIG-COND-RR-HET-HOM-OFF}%
\end{equation}
By contrast, in case of homodyne detection on $\hat{p}$, the corresponding
eigenvalues can be obtained from Eq.~(\ref{SYMP-EIG-COND-RR-HET-HOM-OFF}) by
exchanging $g\longleftrightarrow g^{\prime}$. Averaging over the two
detections, we find the following Holevo bound
\[
\tilde{\chi}_{OFF}^{\blacktriangleleft}=\sum_{k=\pm}\frac{h(\nu_{k})}{2}%
+\frac{1}{2}\log\frac{T(1-T)\mu}{\sqrt[4]{(\omega^{2}-g^{\prime2})(\omega
^{2}-g^{2})}},
\]
which subtracted to the mutual information of Eq.~(\ref{I-OFF-HET-HOM}) gives
the key-rate of Eq.~(\ref{Rate-HET-HOM-RR-OFF}) of the main text, i.e.,
\begin{equation}
\tilde{R}_{OFF}^{\blacktriangleleft}=\frac{1}{2}\log\frac{\sqrt[4]{(\omega
^{2}-g^{2})(\omega^{2}-g^{\prime2})}}{(1-T)\tilde{\Lambda}}-\sum_{k=\pm}%
\frac{h(\nu_{k})}{2}.\label{Rate-HET-HOM-RR-OFF-SI}%
\end{equation}

\end{document}